\def\tc{$T_c$\ }     % critical temperature
\def\bi{$\rm Bi_2Sr_2CaCu_2O_8$\ }  %  Bi:2212
\def\bis{$\rm Bi_2Sr_2Ca_2Cu_3O_{10}$\ }  %  Bi:2223
\def\hr{$H_R$\ }     % resonance field
\def\hrp{$H_R^{\|}$\ }     % resonance field at parallel orientation
\def\hrpp{$H_R^{\bot}$\ }     % resonance field at perpendicular orientation

\documentclass[prb,twocolumn]{revtex4}% Physical Review B
\usepackage{graphicx}% Include figure files
\usepackage{amssymb,amsmath}

\begin{document}

\preprint{APS/123-QED}

\title{Vortex excitations above \tc as revealed by ESR}

\author{Yu. Talanov}
 \altaffiliation{Zavoisky Physical-Technical Institute, 420029, Kazan, Russia.}
\email{talanov@kfti.knc.ru}
\author{L. Salakhutdinov}
\affiliation{Zavoisky Physical-Technical Institute, 420029, Kazan, Russia.}
\author{E. Giannini}
 \affiliation{D\'{e}partement de Physique de la Mati\`{e}re Condens\'{e}e, Universit\'{e} de Gen\`{e}ve,
 1211 Geneva, Switzerland.}
\author{R. Khasanov}
\affiliation{Paul Scherrer Institut, Laboratory for Muon-Spin Spectroscopy,
5232 Villigen, Switzerland.}

\date{\today}

\begin{abstract}
Using electron spin resonance (ESR) technique we have obtained data
evidencing the existence of magnetic vortices in high-temperature
superconductors at temperatures above the critical one $T_c$. We have
studied magnetic excitations in \bis single crystals above \tc with the
method of surface spin decoration. The surface layer of
diphenyl-picrylhydrazyl was used as a sensitive probe of magnetic field
distortions. The temperature dependence of the ESR signal parameters has
indicated that far above \tc the magnetic flux of a sample is affected by
the superconducting order parameter fluctuations while close to \tc its
changes are due to vortex-type excitations.
\end{abstract}

 \maketitle

In recent years the evidences that Cooper pairs are formed at temperatures
above the critical temperature \tc have been obtained with the help of
different methods, such as ARPES \cite{Harris96}, NMR \cite{Zheng00}, Nernst
effect \cite{Pourret06}, resistivity measurements \cite{Raffy07}, tunneling
microscopy \cite{Renner98}, {\em etc}. There are many indications that at
$T>T_c$ the superconducting (SC) order parameter amplitude is not zero,
while the phase coherence is absent. Some scenarios of the phase transition
from the SC state to the normal one imply that upon crossing \tc the phase
coherence destroyed due to the rise of vortex excitations
\cite{Franz07,Tesanovic08}. Thus, it is suggested that the presence of
vortices is the intrinsic property of the pseudogap state of
high-temperature (HT) superconductors. The intensive Nernst signal was
observed in many HTSC materials above \tc \cite{Pourret06,Wang06,Li07}. This
was related to the vortex motion, since it was responsible for the large
Nernst effect in the type-II superconductors. Moreover, the vortex
excitations are manifested in other studies, in particular in the
measurements of the high-frequency conductivity \cite{Corson99}. However,
the vortex existence above \tc is not generally recognized. In the
literature, many different explanations of the large Nernst signal at high
temperatures are proposed, namely, due to the SC fluctuation, without
assuming thermally excited vortices \cite{Serbyn09,Michaeli09}; due to the
unconventional charge density wave \cite{Dora03}; due to the preformed pairs
\cite{Tan04}; due to the interference of the itinerant and localized-carrier
contributions to the thermomagnetic transport \cite{Alexandrov04}.

To elucidate the underlying picture, new experimental data have to be
obtained by means of a method sensitive to the local magnetic perturbations
such as the Abrikosov vortices in the type-II superconductors. In this study
we use electron spin resonance (ESR) of a thin paramagnetic layer
precipitated on a surface of a superconductor (so-called "ESR-decoration")
as such a method. This method was proposed in the work \cite{Rakvin89} to
study the Abrikosov vortex lattice which is formed when a HTSC material
transfers to superconducting state upon lowering the temperature below \tc.
The appearance of vortices results necessarily in the spread of the local
magnetic fields both in the superconductor bulk and on its surface. In
response to the local field dispersion, the ESR signal parameters (the
resonance field value \hr and the line width $\delta H$) of the paramagnetic
substance deposited on the surface are changed. We use this technique to
detect the possible local field perturbations due to the vortex excitations
in the \bis crystals at temperatures above $T_c$.

ESR is a powerful tool to study the local magnetic field distribution of any
origin, in particular that due to the magnetic perturbations produced by the
vortices. Its sensitivity depends on the ESR line width, the narrower the
signal, the higher the resolution. Embedding spin probes in the form of
paramagnetic ions results in a broad signal and lowers considerably the ESR
resolution. To enhance the resolution, the organic free-radical compounds in
the form of a surface layer are used as paramagnetic probes with a narrow
ESR signal. In this study 2,2-diphenyl-1-picrylhydrazyl (DPPH) is used. It
has a narrow Lorentzian ESR signal 1.2\,Oe wide. Its resonance field at the
spectrometer working frequency of 9.3\,GHz is about 3300\,Oe ($g = 2.0036$).

DPPH is deposited on a flat surface of a sample under study. The DPPH layer
of required thickness was obtained by its precipitation from the solution in
benzene. The layer thickness should meet two requirements. On the one hand,
it should not be notably larger than the spatial period of the magnetic
field variation on the sample surface. If the layer is thicker, the ESR
signal is mainly due to DPPH which is not affected by the field
inhomogeneity. The Abrikosov vortex lattice constant in the magnetic field
of 3000\,Oe and $T<T_c$ is estimated as about 80\,nm. So the layer thickness
should not exceed $100\div200$\,nm. On the other hand, the DPPH layer
 less than 100\,nm thick and with the area of several square millimeters does
not provide the ESR signal intensive enough for its analysis. Therefore the
optimal layer thickness, which is sufficiently sensitive to the vortex
perturbations of the magnetic field, has to be $150\div200$\,nm.

ESR spectra were recorded on a Bruker X-band BER-418s spectrometer with
working frequency $9.2\div9.7$\,GHz and magnetic field modulation at a
frequency 100\,kHz. Signal proportional to the first derivatives of absorbed
power was registered with the help of lock-in amplifier EG\&G Model 5209. A
sample was cooled with a helium gas flow inside a Dewar tube passing through
the resonator. Resonance lines were recorded upon sweeping the applied field
up. To increase the accuracy in determining the signal position, a small LiF
crystal containing the dendrites of pure Li metal was mounted in a resonator
along with a sample. Its ESR signal is about 0.1\,Oe wide with $g=2.00226$.

Single crystals of \bis were used because they have the highest \tc
($\sim110$\,K) among all Bi-based HTSC compounds. It allows us to eliminate
the effect of possible impurity phases with \tc higher than that of the main
phase. The single crystals were grown with the travelling solvent floating
zone technique\cite{Giannini04}. Approximate crystal dimensions are
$1.5\times1.5\times0.1$\,mm$^3$.

The superconducting transition was recorded in the measurements of the
temperature dependence of ac-susceptibility at the frequency of 20\,MHz. Its
parameters are $T_c^{onset}=111.6$\,K and the width of 3\,K.

\begin{figure}
\includegraphics[width=7.5cm]{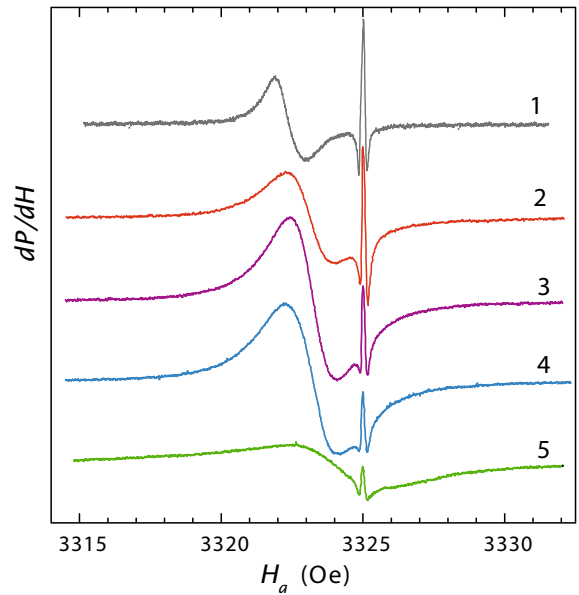}
\caption{\label{spectra} (1) ESR spectrum of the DPPH crystal and the Li
dendrite at $T=290$\,K; (2)-(5) ESR spectra of the DPPH layer and the Li
CESR signal at 232, 125, 108 and 98\,K. $\nu=9317$\,MHz for all spectra.}
\end{figure}

The ESR spectra of the DPPH crystal and the DPPH layer and the conduction
electron spin resonance (CESR) signal of the Li dendrite are shown in
Fig.\ref{spectra}. The spectrum (1) was obtained from the DPPH crystal which
was then dissolved in benzene and deposited on a superconductor. Its signal
has the following parameters: Lorentz shape, peak-to-peak line width $\delta
H_{pp}=1.2$\,Oe, resonance field $H_R=3322.4$\,Oe at $\nu=9317$\,MHz. These
parameters remain unchanged upon lowering the temperature from 300 to 30\,K.
The signal amplitude changes inversely proportional to the temperature. The
CESR signal of the Li dendrite is of the Dyson shape with $\delta
H_{pp}\simeq0.1$\,Oe and $H_R=3325.0$\,Oe. Its line width and resonance
field are temperature-independent, but the shape changes and the amplitude
decreases  with lowering the temperature due to the decrease in the
skin-layer thickness (see Fig.\ref{spectra}, curves (2) - (5)).

\begin{figure}
\includegraphics[width=8.3cm]{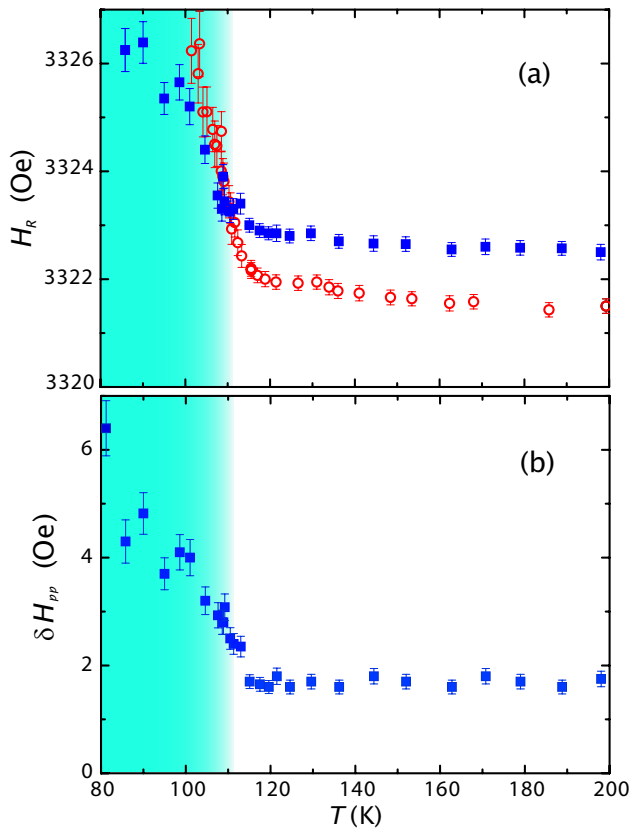}
\caption{\label{Tdepend} (a) Temperature dependence of the resonance field
of the DPPH layer deposited on the \bis crystal for two orientations in the
applied magnetic field: squares - $H_a\|c$, circles - $H_a\|ab$. (b)
Temperature dependence of the line width for the orientation $H_a\|c$.
Shaded region corresponds to the SC state.}
\end{figure}

The behavior of the ESR signal of the DPPH layer on the \bis crystal is
quite different. First of all, ESR signal shifts to higher fields. The shift
depends on the crystal orientation in the applied magnetic field and
temperature (Figs.\ref{spectra} and \ref{Tdepend}). Secondly, the signal
broadens and its broadening depends on temperature as well. The signal
changes are weak far from \tc ($T>120$\,K), but they become observable close
to the critical temperature, particularly below \tc.

\begin{figure}
\includegraphics[width=8cm]{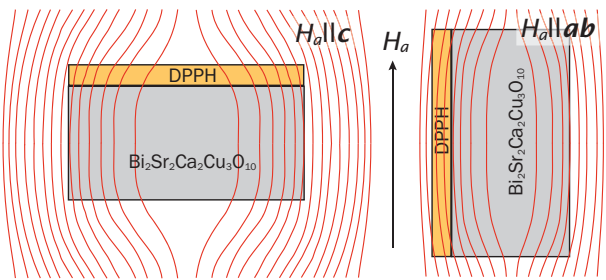}
\caption{\label{flux} Sketch of the magnetic field lines during the flux
penetration into the sample upon increasing the applied field with $H_a\|c$\
and $H_a\|ab$. Field lines are plotted in accordance with
Ref.\cite{Brandt96}}
\end{figure}

The origin of the transformations of the surface-spin-layer EPR signal upon
the SC transition of a HTSC sample was discussed in many publications (see,
for example, Refs.\cite{Rakvin89,Pozek96}). These changes are induced by the
distortions of the magnetic field on the SC surface and are due to three
effects: (i) Meissner shielding leads to the field expulsion from the
superconductor and thus enhances the field value close to it. (ii) The
appearance of the Abrikosov vortex lattice forms a spatially-modulated field
distribution both in the sample bulk and close to its surface outside
\cite{Abrikosov57,Pozek96}. The modulation period is of the distance between
vortices. The mean field value is considerably lower than that of the
applied one. (iii) Any variation of the applied magnetic field results in
the appearance of the vortex density gradients due to vortex pinning on the
crystal structure imperfections. Thus the internal and surface field values
can be both lower or higher than that of the applied one depending on the
field variation direction (up or down, respectively). Figure \ref{flux}
shows the field distribution inside and near an SC strip placed in the
increasing magnetic field which is perpendicular ($H_a\|c$) and parallel
($H_a\|ab$) to the crystal plane surface coated by a DPPH layer. The field
strength and vortex line density are proportional to the density of the
field lines shown in Fig.\ref{flux}.

The above field distortions are reflected in the ESR spectrum of the DPPH
layer. Namely, the decrease in the local field strength results in the
signal shift towards higher fields, and the increase in the local field
shifts the signal towards lower fields. The spatial variation of the field
strength near the SC surface results in the broadening of the ESR line since
different parts of the spin layer are in the fields of different strengths.

The temperature dependence of the resonance field of the DPPH layer is shown
in Fig.\ref{Tdepend}a for two orientations, parallel ($H_a\|ab$) and
perpendicular ($H_a\| c$). As noted above, the \hrp and \hrpp values differ
from the resonance field of the DPPH crystal and from each other. At
$\nu=9315$\,MHz the resonance field of the DPPH crystal is 3321.4\,Oe, while
$H_R^{\|}=3321.7$\,Oe and $H_R^{\bot}=3322.5$\,Oe when temperature is above
150\,K. As the temperature approaches \tc\ from above, the resonance field
increases. Below $T_c$, the upward shift becomes very abrupt,
$\sim0.2$\,Oe/K.

 When the applied field is perpendicular to the SC surface
 ($H_a\|c$ in the present case), the local field distribution is
 determined by the field variations inside the vortex lattice \cite{Rakvin89,Pozek96}
and by the vortex density gradients \cite{Khasanov93}. If the applied field
increases as in the present case, then the internal field in the central
part decreases due to the presence of vortices and their gradients (see
Fig.\ref{flux}). As the temperature is lowered, both contributions enhance
rapidly and result in the larger upward shift. It is shown in many
publications (see, for example, Ref.\cite{Pozek96,Khasanov93}) that at
$T<T_c$ the surface-layer ESR signal shift $\Delta H_R^{\bot}$ is
unambiguously due to the formation of the vortex system and the vortex
density gradients. Since the $H_R^{\bot}(T)$ dependence does not disappear
but extends to the temperatures above $T_c$, one can assume that vortices
exist in this temperature range.

A small but noticeable signal shift at $T>140$\,K has very a weak or even
zero temperature dependence and is likely of another origin. The direction
and value of the shift indicate the presence of the constant diamagnetic
contribution to the magnetic field on the sample surface. This contribution
correlates well with the magnetization measurements of the \bis crystals
\cite{Li93}. It has been found at $H_a\| c$\ and from 110 to 120\,K the
$4\pi M$ magnitude was about 1\,G. The authors \cite{Li93} attributed this
magnetization to the Ginzburg-Landau fluctuation effect. There are no
magnetization data on \bis at $H_a\| ab$ in the literature. For the analysis
we can use the susceptibility measurement results obtained for both
orientation of the \bi crystal in the magnetic field \cite{Pomar96}. The
authors \cite{Pomar96} found a noticeable contribution to the magnetic
susceptibility ($\Delta\chi\sim10^{-5}$) in the large temperature range from
\tc to 300\,K. This contribution, which was attributed to the thermal
fluctuations of the SC order parameter amplitude, is not temperature-depend
but varies with orientation. The difference between $\chi_{ab}$\ and
$\chi_c$\ is about $3\cdot10^{-5}$, which corresponds to $4\pi
M\simeq0.1$\,G. It is markedly smaller but comparable with the resonance
field difference in our ESR measurements from 120 to 200\,K: $\Delta
H_R\simeq0.9$\,Oe. The disagreement can be due to both by the differing
compound and by the different oxygen doping level. Thus the assumption that
the ESR signal shift to higher fields at $T>120$\,K is due to the SC
fluctuation contribution to the magnetization seems to be quite reasonable.

The $H_R(T)$ dependence (both parallel and perpendicular) is similar to the
$M(T)$\ dependence in Ref.\cite{Li93}. However, the authors \cite{Li93}
connected the abrupt increase in the diamagnetic magnetization close to \tc
(above and below) with the increased contribution of the SC order parameter
fluctuations. It contradicts with our explanation of the resonance field
behavior by the presence of vortices. The vortex character of the magnetic
flux distortion is unduobtful since the presence of vortices (at least just
below $T_c$) is experimentally proved by many methods of the vortex
visualization, in particular by Bitter decoration \cite{Bolle91}, magnetic
force microscopy \cite{Moser95}, Lorentz microscopy \cite{Harada93},
scanning Hall probe microscopy \cite{Oral97}, scanning SQUID microscopy
\cite{Iguchi01} {\em etc}. Therefore their existence and the effect on
magnetization below \tc are beyond question. The fact, that the $H_R(T)$\
function is continuous and monotonic at crossing \tc, suggests that below
and above critical temperature the field distortion has the same origin -
vortex excitations.

It should be noted that in the strict sense the vortex effect discussed
above is only valid for $H_a\| c$. At $H_a\| ab$ the vortex effect should be
different (see Fig.\ref{flux}). Namely, the magnetic field expulsion due to
the Meissner shielding results in the increase in the magnetic field
strength on the surface spin-layer. The ESR signal should move to lower
fields respectively. However, the observed shift is opposite. This
observation can be explained taking into account the following. First, since
the crystal thickness is by order of magnitude less than its plane
dimension, the demagnetization factor is very small, $N_{\|}\ll1$. So the
field increase near the surface and the relative signal shift can not be
considerable \cite{Pencarina95}. Second, in the vortex state some of the
vortex lines bend near the edges and run both through the surface
perpendicular to applied field and through parallel surface with DPPH
deposited on it (Fig.\ref{flux}). The last effect has to result in weakening
field strength with the corresponding ESR signal shift to higher fields. The
signal shift below and slightly above \tc suggests just vortex contribution
to field distribution on the examined surface of the \bis crystal. The
temperature dependence of the ESR line width of the DPPH layer
(Fig.\ref{Tdepend}b), which is discussed below, suggests the same.

Note, that for both $H_a\| ab$ and $H_a\| c$ the $\delta H_{pp}(T)$
dependence is similar. The only difference is in its slope below $T_c$. Upon
lowering temperature the line width is constant within the experimental
error from 300\,K down to $T'\simeq115$\,K. An additional line broadening
appears below $T'$\ but above \tc, it increases with lowering temperature
and is most pronounced when the sample transfers to the SC state. Taking
into account the resonance field behavior (Fig.\ref{Tdepend}a), the ESR
signal broadening can be due to the appearance and development of the vortex
structure as well. The $\delta H_{pp}(T)$\ dependence reveals the onset of
the additional broadening due to the vortex excitation by several degrees
above $T_c$. The constant signal shift observed at higher temperatures is
obviously due to the contribution from the superconducting order amplitude
fluctuations which are uniformly distributed over the sample or averaged
rapidly and do not result in the resonance line broadening. However, there
is an experimental indication of the possibility of the nonuniform
distribution of such fluctuations. In Ref.\cite{Iguchi01} the diamagnetic
regions larger than $10\,\mu$m were revealed by the scanning SQUID
microscopy technique in the $\rm La_{2-x}Sr_xCuO_4$ thin films in the
temperature range from \tc to about $2T_c$. Such diamagnetic regions can be
formed in \bis crystals as well and produce the spin-layer signal broadening
in the wide temperature range above $T_c$. But this is not observed. The
situation can be explained by estimating the additional field due to the
diamagnetic regions. According to Ref.\cite{Iguchi01} the $B_{dia}^{loc}$\
value near \tc is about $5\,\mu$T\ (0.05\,G). It makes only 2.5\,\% of the
line width, that is within the experimental error. When the temperature is
far above \tc ($T\simeq T_c+40$\,K), the addition $B_{dia}^{loc}$\ becomes 5
times larger, but the fractional area of the diamagnetic regions at this
temperature is one order of magnitude less than the total
area\cite{Iguchi01}. Such diamagnetic contribution is not detectable with
our technique.

In conclusion, the surface-probe ESR study of the features of the magnetic
state of the \bis single crystal in the vicinity of a critical temperature
reveals the presence of the magnetic flux disturbances on the crystal
surface at $T>T_c$. The temperature dependence of the ESR signal shift and
line width at $T\gtrsim T_c$\ is the same as that observed upon the
formation of the vortex system in the superconducting material. This
supports the hypothesis that the vortex excitations exist in the normal
state of a superconductor, that is, above $T_c$.\bigskip

We thank G. B. Teitel'baum and G. G. Khaliullin for helpful discussions. The
study was supported by the Russian Foundation for Basic Research under Grant
No. 07-02-01184-a.

\bibliography{biblio}
\end{document}